# Swift heavy ion irradiation induced structural phase transitions in $BaTiO_3$: An *in-situ* x-ray diffraction study

Ravindra Singh Solanki[1*], Jay Prakash Patel[1], Chandana Rath[1], Pawan Kumar Kulriya[2], Devesh Kumar Avasthi[2], Dhananjai Pandey[1**]

[1]School of Materials Science and Technology, Indian Institute of Technology (Banaras Hindu University), Varanasi-221005, India

[2]Inter-University Accelerator Centre, Aruna Asaf Ali Marg, New Delhi-110067, India

## Abstract

There is considerable controversy about swift heavy ion (SHI) irradiation induced displacive phase transitions in thermally insulating oxides. We present here unambiguous evidence for tetragonal to monoclinic and rhombohedral to monoclinic phase transitions in $BaTiO_3$ under swift heavy ion irradiation (120MeV $^{108}Ag^{+9}$ ions) using *in-situ* x-ray powder diffraction (XRPD) studies. The anomalous splitting/broadening of $111/222_{pc}$, $200_{pc}$ and $220_{pc}$ pseudocubic peaks for fluences $\geq 3\times10^{12}$ ions/cm$^2$ reveal structural changes before amorphization at higher fluences. Lebail analysis of XRPD profiles confirm that the monoclinic phase is of $M_A$ type in the Cm space group. Shear stress for the structural phase transition is estimated to be ~ 430MPa, which we believe is generated as a result of stopping of the SHI.

*Present address: Centre of Material Sciences, Institute of Interdisciplinary Studies, University of Allahabad, Allahabad-211002, Uttar Pradesh, India.

**Corresponding author's email: dp.mst1979@gmail.com

Passage of swift heavy ions in the metallic materials is known to induce several interesting phenomena such as dramatic growth perpendicular to the ion beam without any volume change and mass transport in glassy $Pb_{80}Si_{20}$ [1], amorphisation of crystalline $Ni_3B$ ribbons [2], grain alignment and rotation in nanocrystalline Ti layers [3], growth of Au nanoparticles embedded in a silica film [4], creep of crystalline metallic layer [5], formation of quasi one-dimensional magnetic nanostructures in inter metallic rare earth-transition metal compounds [6], structural changes [7] and amorphisation of crystalline phases [8]. In thermally insulating oxides, structural phase transitions from monoclinic to tetragonal in $ZrO_2$ [9, 10] and anatase to rutile in $TiO_2$ [11] induced by 300 MeV Ge-ion irradiation and 200MeV Ag ions, respectively, have been reported. In the same work [10], 300MeV Ni ion irradiation could not induce any phase transition in $ZrO_2$ and this was attributed to the electronic energy loss required for phase transition from monoclinic to tetragonal was less than ~ 12 keV/nm in 300 MeV Ni ion [9, 10]. Irradiation of yttria stabilized cubic zirconia with 940MeV Pb ions with an electronic energy loss of 44.3 keV/nm leads to broadening of diffraction peaks due to radiation induced defects without any phase transition [12]. It has also been pointed out that the ion irradiation could only generate hydrostatic stress field and causes significant atomic arrangements without any structural phase transition [12]. *In case of $BaTiO_3$, 1 MeV Au and 300keV Ne ion irradiation has been reported to lead to amorphisation [13, 14]. Thus, there are conflicting reports about the structural phase transitions being induced by swift heavy ion (SHI) irradiation in oxides.* In this letter, we present unambiguous experimental evidence for tetragonal to monoclinic and rhombohedral to monoclinic phase transitions in $BaTiO_3$ (BT) ceramics induced by 120MeV $^{108}Ag^{+9}$ ions (electronic energy loss= 8.0 keV/nm) with varying fluences at 300K and 96K, respectively, using *in-situ* x-ray powder diffraction studies. The shear stress required for inducing such structural phase transitions is estimated to be ~430 MPa. *The observation of monoclinic phase in* $BaTiO_3$ *as a result of SHI irradiation acquires special significance for its application as a piezoelectric ceramic as the high piezoelectric response of the well known Pb-based piezoelectric ceramics, like* $Pb(Zr_xTi_{1-x})O_3$ *(PZT) and* $(1-x)Pb(Mg_{1/3}Nb_{2/3})O_3$-$xPbTiO_3$ *(PMN-xPT,) is known to be linked with compositions showing monoclinic structures [15, 16].*

*The details of sample preparation, SHI irradiation, in-situ x-ray diffraction and analysis of data are given in the supplementary file [17].* The structure of the perovskite form of BT in the paraelectric state is cubic in the $Pm\bar{3}m$ space group. It undergoes a succession of ferroelectric phase transitions below 403, 278 and 183 K leading to tetragonal, orthorhombic and rhombohedral phases in the P4mm, Amm2 and R3m space groups, respectively. Locally, all these phases of BT are known to be rhombohedrally distorted [18, 19]. Fig. 1(A) depicts the evolution of XRD profiles (after stripping off $K_{\alpha_2}$) of $111_{pc}$, $200_{pc}$ *and* $220_{pc}$ (pc stands for pseudocubic) perovskite reflections of a pristine and ion beam irradiated BT pellet at room temperature with varying ion fluences. At zero fluence (pristine sample), the $111_{pc}$ perovskite peak is a singlet while the $200_{pc}$ and $220_{pc}$ peaks are doublet. This is the characteristics of the tetragonal phase of BT. However, the intensity of the $002_{pc}$ peak is more than that of $200_{pc}$ peak which is due to texture effect in the pellet. As the ion fluence increases, the broadening/splitting of peaks increases. Interestingly, for ion fluence of $3x10^{12}$ ions/cm$^2$, the $111_{pc}$ peak splits into two and at the same time the $200_{pc}$ and $220_{pc}$ peaks remain doublet albeit much broadened. The broadening of the peaks with increasing ion fluences is essentially due to radiation induced defects which distort the lattice [20]. The decrease in intensity of the peaks in the x-ray

diffraction pattern is due to the presence of the amorphized columnar tracks in irradiated sample [21]. Electronic stopping is mainly responsible for decrease of intensity of XRD peaks with increasing fluence of heavy ion [20, 21]. The splitting of the $111_{pc}$ peak of the tetragonal phase of BT suggests that it undergoes a structural phase transition to a new phase. To determine the true symmetry of this irradiation induced phase, we have carried out Lebail analysis of *in-situ* XRD patterns and the results are presented in Fig. 2(A). Lebail refinement for the pristine sample using the P4mm space group model gives satisfactory fits for all the peaks, as expected. On the other hand, Lebail fits for the same BT pellet after irradiation with $6\times10^{12}$ ions/cm$^2$ fluence using P4mm space group model is not able to account for the anomalous broadening nature of $200_{pc}$ reflection and also the splitting of the $111_{pc}$ and $220_{pc}$ peaks as can be seen from Fig. 2A(b). This indicates that other lower symmetry space groups have to be considered to explain the observed XRD profiles. *On increasing the ion fluence beyond 1x10$^{12}$, the broadening of the XRD peaks becomes unusually large, as can be seen from Fig. 1(a). This is a signature of amorphisation of the sample volume being probed by the XRD.*

The ISOTROPY software package [22] can be used to determine the sequence of phase transitions in terms of irreducible representations (IR) of the modes of the cubic perovskite phase. The freezing of the different $\Gamma_4^-$ irrep (k=0, 0, 0) (zone centre optical phonon) of the cubic $Pm\bar{3}m$ phase can lead to different phases with different space group symmetries: (a) tetragonal, space group: P4mm, (b) orthorhombic, space group: Amm2, (c) rhombohedral, space group: R3m, (d) monoclinic, space group: Pm, (e) monoclinic, space group: Cm and (f) triclinic, space group: P1. Lebail fits corresponding to the structural models other than triclinic are shown in Fig. 2A (c to f). As already discussed, the P4mm space group is not able to account for the splitting of $111_{pc}$ and $220_{pc}$ peaks. The Amm2 and R3m space group models give better fit to the $111_{pc}$ reflection but they are not able to account for the peak position of $220_{pc}$ reflections. It is evident from Fig. 2A(c) and (d) that there is a huge mismatch between the observed and calculated peak positions. We therefore discard these two space group models as well. Figs. 2A(e) and (f) depict the Lebail fits using Pm and Cm space group models. These space group models give nearly comparable fit, but as evident from a comparison of these figures, the Cm space group model gives better fit to the $111_{pc}$ and $200_{pc}$ perovskite reflections with a lower value of $\chi^2$ and $R_{wp}$. Our analysis thus shows that the monoclinic phase in the Cm space group represents the true symmetry of BT after ion irradiation with a fluence $>3\times10^{12}$ ions/cm$^2$ but before amorphization. The equivalent pseudocubic lattice parameters ($a_p$, $b_p$, $c_p$) of the monoclinic phase were calculated using the relationship $a_p \simeq a_m/\sqrt{2}$, $b_p \simeq b_m/\sqrt{2}$ and $c_p \simeq c_m$, where $a_m$, $b_m$, $c_m$ are the monoclinic cell parameters. These are plotted as a function of ion-fluence in Fig. 3(a). For fluence of $3\times10^{12}$ ions/cm$^2$, the lattice parameters of monoclinic phase in terms of pseudocubic cell are $a_p$=4.04193 Å, $b_p$=4.02426 Å, and $c_p$=4.1436 Å. These lattice parameters show pseudotetragonal ($a_p \approx b_p < c_p$) relationship, implying $P_x \simeq P_y < P_z$. This phase is therefore of the $M_A$ type in the notation of Vanderbilt and Cohen [23] and is identical to that reported below room temperature for $Pb(Zr_xTi_{1-x})O_3$ (PZT) with x=0.520 [24, 25]. If we extrapolate the pseudocubic lattice parameters, it is found that they all meet for a fluence of ~$2.55\times10^{13}$ ions/cm$^2$. This suggests that the tetragonal phase of BT should eventually transform to the rhombohedral phase through the intermediate monoclinic phase of $M_A$ type. However, due to ion irradiation induced defects, it gets amorphised at much lower fluences before transforming to the rhombohedral phase.

We now proceed to discuss the effect of ion-beam irradiation on the rhombohedral phase of BT recorded at 96 K. Fig. 1(B) depicts *in-situ* XRD patterns ($K_{\alpha 2}$ stripped) of BT pellet recorded at 96 K with varying ion fluences. It is evident from Fig. 1(B) that the in-situ XRD patterns for pristine BT sample recorded at 96 K, the $200_{pc}$ peak is a singlet while $222_{pc}$ peak is a doublet. These are the characteristics of the rhombohedral phase. We therefore refine the XRD pattern using R3m space group stable at T < 183 K in unirradiated sample. It is evident from the fits shown in Fig. 2B(a) that all the peaks can be accounted for using the R3m space group satisfactorily. Thus, our pristine BT pellet at 96 K is rhombohedral in the R3m space group. As, we increase the fluence, the broadening of all the peaks starts increasing and for an ion fluence of $3\times10^{12}$ ions/cm$^2$, all the representative perovskite reflections become doublets and much broader. *For much higher fluences ($\gtrsim 1\times10^{13}$ ions/cm$^2$), the broadening/splitting disappears and intensity of all the peaks decreases drastically. This indicates the amorphization of the BT pellet. Thus, there is a phase transition from rhombohedral to a new phase for fluence $>1\times10^{12}$ ions/cm$^2$ but before amorphization.* This new phase according to the splitting/broadening of peaks may also be a monoclinic phase. To determine the space group symmetry of the phase stable for fluence $=3\times10^{12}$ ions/cm$^2$, we show in Fig. 2(B) the Lebail fits for the various space group models obtained using ISOTROPY software package [22] as discussed above in relation to the tetragonal phase of BT. We first refined the XRD pattern using rhombohedral R3m space group which is the stable phase of BT pellet at 96 K. However, due to the doublet nature of $200_{pc}$ peak, this space group model is not able to account for this peak. Further, this model does not give good fit to the $220_{pc}$ reflection. Thus, this model does not represent true symmetry of the phase resulting after irradiation of the R3m phase. Thereafter, we considered Amm2 and P4mm space group models. As evident from Fig. 2B(c) and (d) that these space group models improve the fits to the $200_{pc}$ peak as well as other peaks. But the Pm and Cm space group models for which refinements are presented in Fig. 2B(e) and (f) give even better fits with lower values of $\chi^2$ and $R_{wp}$. *A comparison of the fits between Pm and Cm space group models shows that Cm space group model gives a little better fit and also lower values of* $\chi^2$ *and* $R_{wp}$. We therefore propose that Cm space group represents the true structure of the phase after irradiation at fluence of $\sim 3\times10^{12}$ ions/cm$^2$ and above but before amorphization. Fig. 3(b) depicts the variation of the pseudocubic lattice parameters with increasing ion-fluence. As the irradiation dose increases, $a_p$ and $c_p$ decreases for rhombohedral phase but for the monoclinic phase pseudocubic lattice parameters increase. The refined lattice parameters for fluence of $3\times10^{12}$ ions/cm$^2$ are $a_p$=3.9977 Å, $b_p$=4.0035 Å and $c_p$=4.0451 Å. Thus, they show pseudotetragonal ($a_p \approx b_p < c_p$) relationship. This phase is therefore also of $M_A$ type as reported for PZT [24, 25].

The high piezoelectric response in Pb-based piezoelectric ceramics like PZT have attributed to the existence of monoclinic phases (see reviews by Noheda et al. [15] and Pandey et al. [16]) because their spontaneous polarization direction can rotate on a crystallographic symmetry plane unlike those of the tetragonal, orthorhombic and rhombohedral phases where it is constrained to lie along a specific crystallographic direction. In tetragonal and rhombohedral phases, the polarization vector is along $[001]_{pc}$ and $[111]_{pc}$ directions, respectively, while it lies in the $(1\bar{1}0)_{pc}$ plane for the $M_A$ type monoclinic Cm phase as depicted schematically in Fig.3(c). The monoclinic Cm space group is a subgroup of both P4mm and R3m phases and can therefore act as a bridging phase [15]. In fact, the tetragonal P4mm phase can transform to the rhombohedral R3m phase through the bridging monoclinic $M_A$ phase by rotating the polarization vector from [001] to [111] on the $(1\bar{1}0)_{pc}$ plane of the monoclinic $M_A$ phase. Thus, the room temperature tetragonal phase of $BaTiO_3$ which transforms to the monoclinic $M_A$ phase on increasing the ion fluence should have finally transformed to the rhombohedral phase. However, due to the increased defects in BT pellets at high fluence levels, the $M_A$ phase seems to amorphize before transforming to the rhombohedral R3m phase. Same is the case for the rhombohedral phase, where ion beam irradiation transforms the rhombohedral phase to the $M_A$ type monoclinic phase but the target eventually amorphizes before becoming pure tetragonal. While tetragonal to monoclinic phase transition has been reported in $BaTiO_3$ under external electric field [26, 27] and epitaxial constraints [28], our results show such transition can also be induced by SHI irradiation which has not been reported earlier.

The tetragonal structure may transform to $M_A$ type monoclinic phase by applying a shear stress on the $(001)_{pc}$ plane of the tetragonal phase [24]. We have estimated the shear stress required for this transition using the formula $\sigma_{ij} = c_{ijkl}\, \epsilon_{kl}$, where $c_{ijkl}$ are the elastic coefficients in the tetragonal state below the Curie point and $\sigma_{ij}$ and $\epsilon_{kl}$ are stress and strain tensors, respectively [29]. For the tetragonal phase of BT, the value of shear elastic coefficient ($c_{1313}$) is = $6.2\times10^{10}$ N/m$^2$ [29] and for the tetragonal to monoclinic phase transition, the shear angle, as obtained from the Lebail refinement, is 0.398 (0.00694 radians). The calculated transformation stress is $\sigma_{ij}$ = $0.043\times10^{10}$ N/m$^2$ or 430 MPa. Thus, ion-beam radiation creates a shear stress of 430 MPa to transform the tetragonal phase of BT into the monoclinic phase. In pure zirconia, a phase transition from monoclinic to tetragonal phase [9, 10] has been reported by 300 MeV Ge ion at $1.2\times10^{13}$ ions/cm$^2$ fluence. This transition also requires shear stresses and ion beam bombardment presumably generates such shear stresses. A similar type of shear stress generation due to ion beam bombardment has been reported in a $Pd_{80}Si_{20}$ metallic glass where it leads to anisotropic plastic deformation [30].

The energetic ions on passing through a solid target slow down as a result of nuclear and electronic stopping, the latter dominating at high ion energies such as those used in the present investigation [20, 21]. *Passage of SHI is known to cause displacive type structural phase transitions involving cooperative displacement of atoms in metallic and alloy targets [see e.g. 31] due to strong electron-phonon coupling and good thermal conductivity. This coupling has been shown to be most efficient in metallic systems showing a soft phonon mode that drives the structural phase transition [32]. However, the situation in oxides is quite different because of their low thermal conductivity and much weaker electron-phonon interaction. This is expected to lead to a more pronounced damage and amorphization rather than cooperative displacement of atoms required for a displacive structural phase transition. The observation of structural phase transitions in oxides like BT in the present work and in zirconia ($ZrO_2$) and hafnia ($HfO_2$), as reported earlier, due to SHI irradiation is therefore anomalous. This anomalous behavior in $ZrO_2$ and $HfO_2$ has been explained in terms of a double ion impact mechanism [10] in which the first ion impact creates 'shear-sites' in the form of radiation defects while the second ion generates the shear stresses required for the structural phase transition. The fact that all the peaks in Fig.1 broaden as a result of ion irradiation suggests that radiation defects are indeed getting formed in BT also during the first ion impact. Some of these sites may also get hit by a second ion that may lead to the generation of shear stresses for the transformation of the unamorphized adjoining region to the monoclinic phase as per the double-ion impact mechanism [10]. Since the number of radiation defect* *sites are growing continuously with increasing fluence, the material at high enough fluence would eventually become amorphous. This explanation requires theoretical calculations in a future work to understand the energy transfer mechanisms during SHI irradiation in a poor thermal conductor like* $BaTiO_3$.

To summarize, we have shown that the tetragonal and rhombohedral phases of $BaTiO_3$ stable at room temperature and below *183 K* transform to a monoclinic phase in the Cm space group on irradiation with 120MeV Ag swift heavy ions with varying fluences. The irradiation induced monoclinic phase is of $M_A$ type discovered recently in PZT. Although such types of monoclinic phases were reported in thin films *[28]* and single crystals of BT *[26, 27]* induced by epitaxial strain and electric field, respectively, no such phase was observed so far in bulk BT polycrystalline samples under ion irradiation. The estimated shear stress required for the observed tetragonal to monoclinic transition is ~ 430 MPa which we believe to have been generated during SHI irradiation.

We acknowledge the Director, Inter University Accelerator Centre (IUAC), New Delhi for extending ion beam facilities to us. D. Pandey acknowledges financial support from Science and Engineering Research Board (SERB) of India through the award of J. C. Bose National Fellowship grant. R. S. Solanki acknowledges financial support from SERB, India in the form of Research Assistantship under J. C. Bose Fellowship. D. K. Avasthi acknowledges the financial support from DST to setup an in-situ x-ray diffractometer in beam line under the program of Intensification of Research in High Priority Areas (IRHPA).

## Figure Captions

**Fig. 1.** (Color online) *In-situ* x-ray diffraction profiles of $111/222_{pc}$, $200_{pc}$ and $220_{pc}$ peaks of $BaTiO_3$ (A) recorded at room temperature and (B) at 96 K with increasing ion-fluences.

**Fig. 2**. (Color online) Observed (dots), calculated (continuous line) and difference (bottom) XRD profiles for $111/222_{pc}$, $200_{pc}$ and $220_{pc}$ perovskite reflections after Lebail refinement of the structure (A) at room temperature and (B) at 96 K for pristine and irradiated pellet of BT with ion fluence $6\times10^{12}$ ions/cm$^2$, using (a) P4mm, (b) P4mm, (c) Amm2, (d) R3m, (e) Pm and (f) Cm space groups.

**Fig. 3**. (Color online) (a) Variation of tetragonal ($a_t$ and $c_t$) and pseudocubic lattice parameters ($a_p$, $b_p$ and $c_p$) and monoclinic angle β of the monoclinic phase as a function of ion-fluences, (b) variation of pseudocubic lattice parameters for the rhombohedral R3m and monoclinic Cm phases as a function of ion-fluences and (c) schematic depiction of the rotation of the polarization vectors [001] and [111] of the tetragonal P4mm (T) and rhombohedral R3m (R) phases, respectively, leading to the emergence of the monoclinic Cm phase of $M_A$ type.

**Fig. 1**

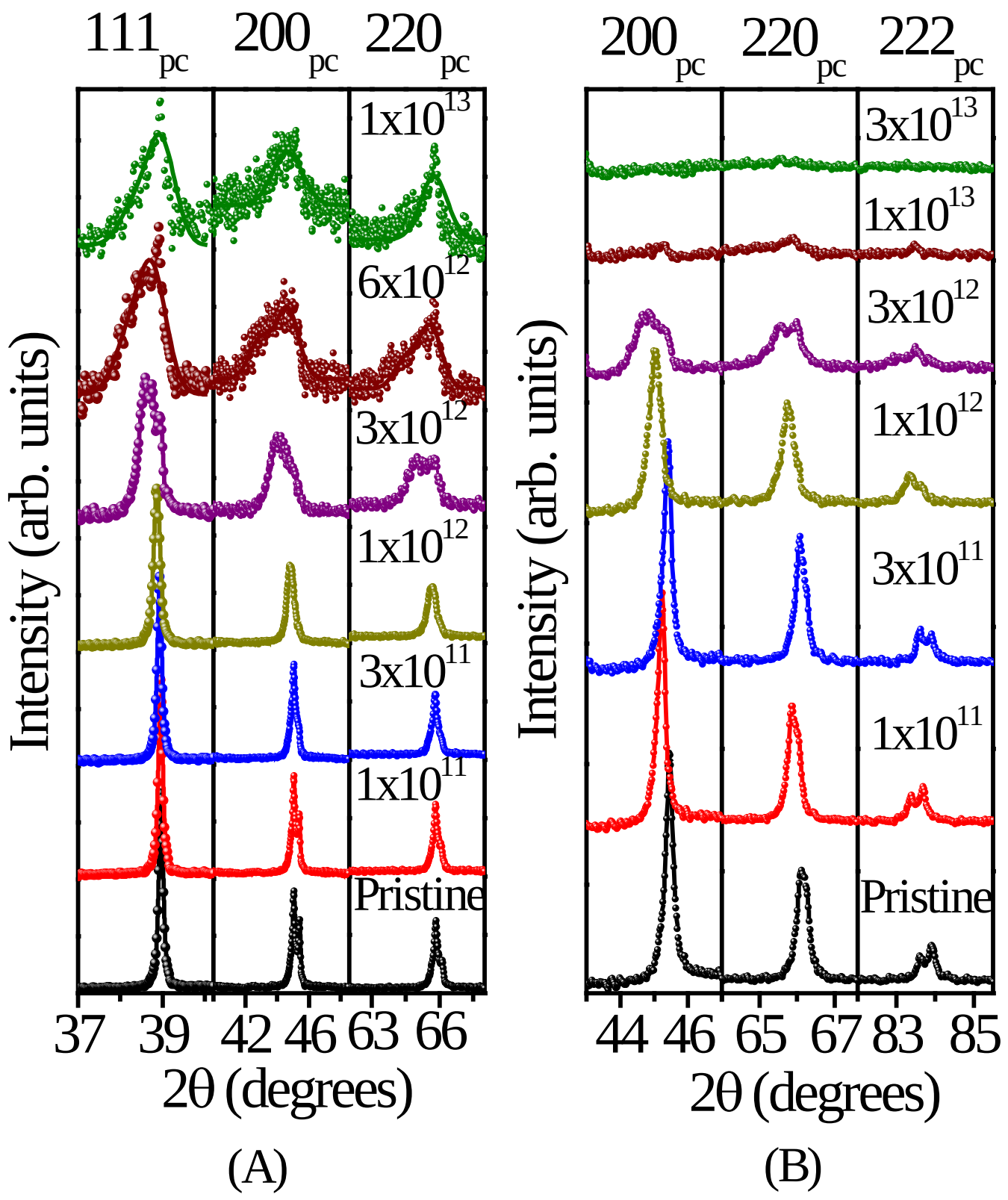

Fig. 2

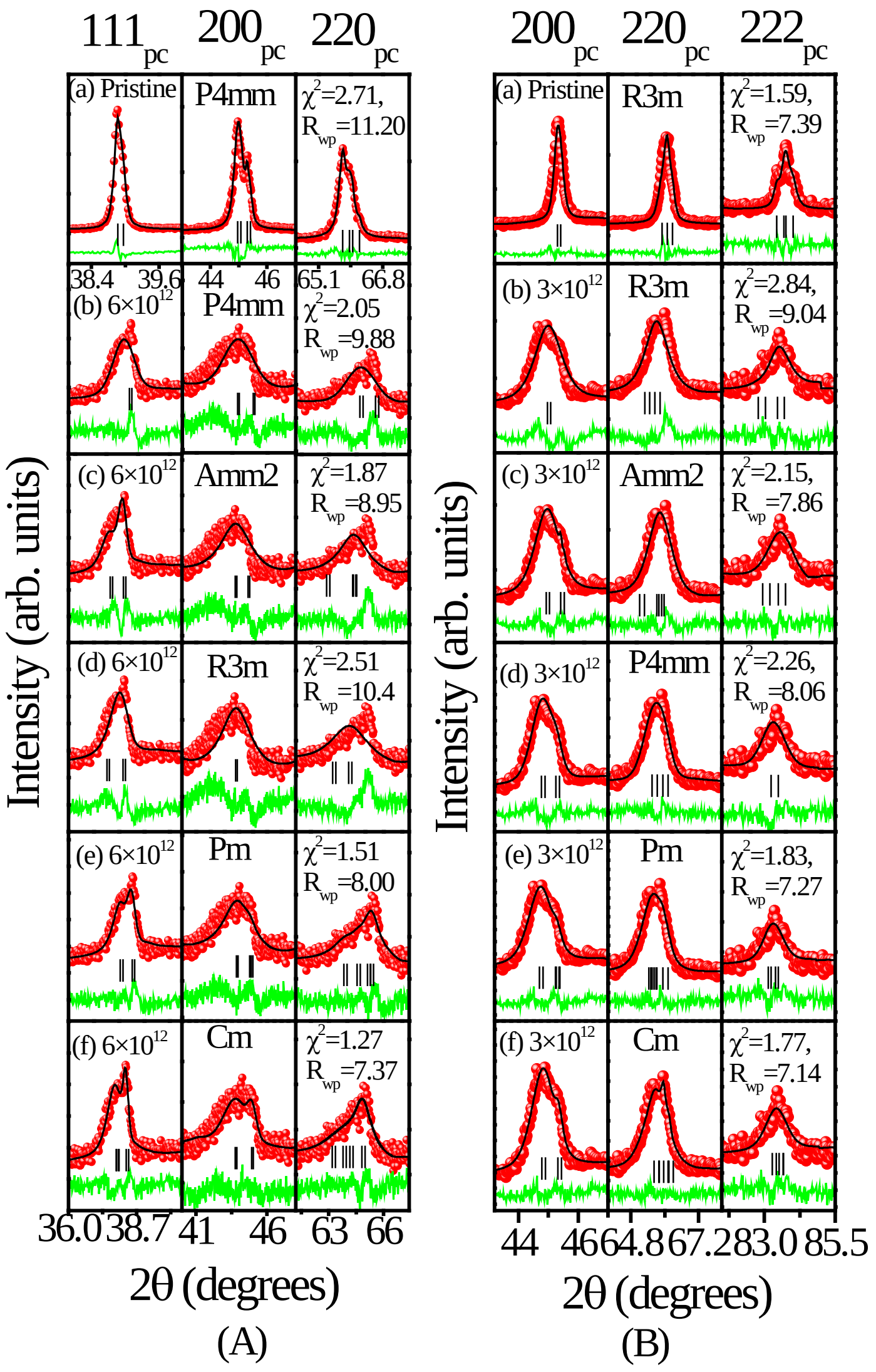

Fig. 3

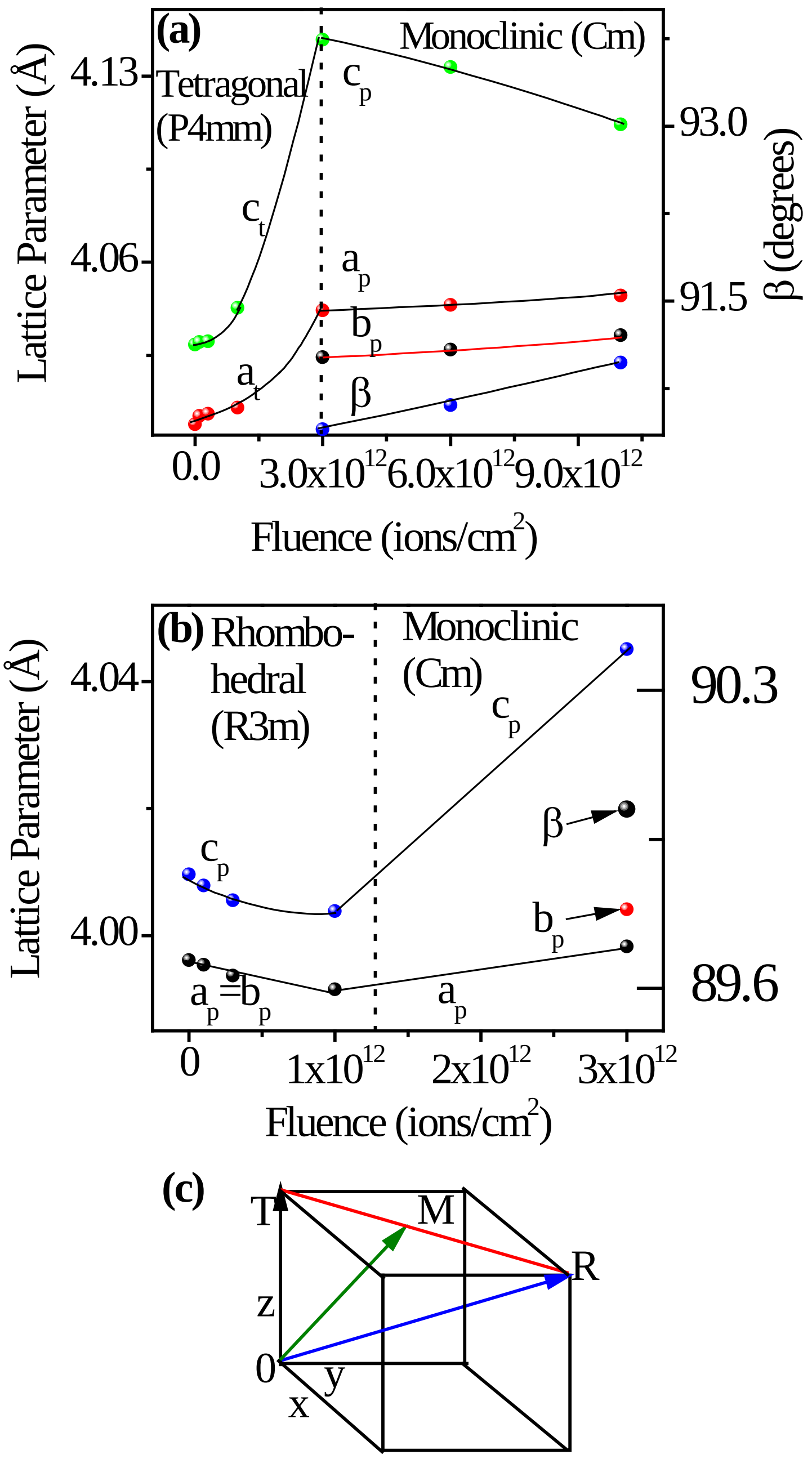

## Supplementary Information

## Experimental details and Analysis

$BaTiO_3$ (BT) powder samples were prepared by conventional solid solution route using analytic reagent (AR) grade $BaCO_3$ and $TiO_2$ with a minimum assay of 99 %. Stoichiometric mixture of $BaCO_3$ and $TiO_2$ was ball milled using zirconia balls and jars and calcined at 1100 °C for 6 h. The calcined powder was compacted in the form of circular pellets in a cylindrical die using a uniaxial hydraulic press at an optimized load of 65 kN. 2 % polyvinyl alcohol (PVA) solution in water was used as a binder. Pellets were sintered at 1350 °C for 6 h in open atmosphere. *The polycrystalline sintered sample of BT with a thickness of ~1mm was irradiated at the Materials Science Beamline, Phase-II of the Pelletron accelerator at the Inter University Accelerator Centre, New Delhi, India.* Vacuum inside the XRD chamber was maintained with the help of a rotary and turbo molecular pump and it was ~$10^{-5}$ torr during irradiation. The irradiation was done using $^{108}Ag^{+9}$ ions with a kinetic energy of 120 MeV. This energy corresponds to nuclear and electronic stopping power of 0.04 keV/nm and 8.0 keV/nm, respectively as calculated using SRIM software [1]. In the present case, the electronic stopping power dominates over the nuclear stopping power, but, still, each ion generates about 45890 vacancies/ions, as calculated form TRIM calculation [1]. The ion fluence was varied from $1\times10^{11}$ to $3\times10^{13}$ ions/cm$^2$. The in-situ x-ray diffraction patterns were recorded from 10° to 90° at a step of 0.02 in $2\theta$ in Bragg-Brentano geometry *at a wavelength of 1.54* Å *(~ 8 keV).* The high speed position sensitive Vantec detector [2] was used for recording of XRD patterns. Lebail refinement of the XRD data was carried out using FULLPROF software package [3]. *The stopping range of 120MeV Ag ion, as calculated using SRIM program [1], in $BaTiO_3$ is ~10* μm *whereas the penetration depth for ~8keV, X-rays is ~7* μm *suggesting that nearly entire irradiated volume is being probed by the X-rays.*